\pdfoutput=1
\documentclass{sig-alternate}
\doi{}
\isbn{}

\usepackage{colortbl}
\usepackage{graphicx}
\usepackage{epsfig}
\usepackage{amssymb}
\usepackage{amsmath}
\usepackage{array}
\usepackage{algorithm}
\usepackage[noend]{algpseudocode}
\usepackage{bm}
\usepackage{bbm}
\usepackage{multirow}
\usepackage{tabularx}
\usepackage{esvect}
\usepackage{adjustbox}
\usepackage{booktabs}
\usepackage{subfigure}
\usepackage[cmyk]{xcolor}

\begin{document}

\title{Multi-source Hierarchical Prediction Consolidation}
\numberofauthors{6}
\author{
 Chenwei Zhang$^\dag$
 \mbox{\hspace{.1in}}
 Sihong Xie$^\sharp$
 \mbox{\hspace{.1in}}
 Yaliang Li$^\ddag$
 \mbox{\hspace{.1in}}
 Jing Gao$^\ddag$
 \mbox{\hspace{.1in}}
 Wei Fan$^\natural$
 \mbox{\hspace{.1in}}
 Philip S. Yu$^{\dag\S}$\\
 \affaddr $^\dag$Department of Computer Science, University of Illinois at Chicago, Chicago, IL, USA\\
 \affaddr $^\sharp$Computer Science and Engineering Department, Lehigh University, Bethlehem, PA, USA\\
  \affaddr $^\ddag$SUNY Buffalo, Buffalo, NY, USA\\
  \affaddr $^\natural$Baidu Research Big Data Lab, Sunnyvale, CA, USA\\
  \affaddr $^\S$Institute for Data Science, Tsinghua University, Beijing, China\\
  \scalebox{.9}{\email{$^\dag$\{czhang99,psyu\}@uic.edu, $^\sharp$sxie@cse.lehigh.edu, $^\ddag$\{yaliangl,jing\}@buffalo.edu, $^\natural$fanwei03@baidu.com}}
}

\maketitle
\begin{abstract}
In big data applications such as healthcare data mining, due to privacy concerns, it is necessary to collect predictions from multiple information sources for the same instance, with raw features being discarded or withheld when aggregating multiple predictions. Besides, crowd-sourced labels need to be aggregated to estimate the ground truth of the data. Because of the imperfect predictive models or human crowdsourcing workers, noisy and conflicting information is ubiquitous and inevitable. Although state-of-the-art aggregation methods have been proposed to handle label spaces with flat structures, as the label space is becoming more and more complicated, aggregation under a label hierarchical structure becomes necessary but has been largely ignored. These label hierarchies can be quite informative as they are usually created by domain experts to make sense of highly complex label correlations for many real-world cases like protein functionality interactions or disease relationships.

We propose a novel multi-source hierarchical prediction consolidation method to effectively exploits the complicated hierarchical label structures to resolve the noisy and conflicting information that inherently originates from multiple imperfect sources. We formulate the problem as an optimization problem with a closed-form solution. The proposed method captures the smoothness over all information sources as well as penalizing any consolidation result that violates the constraints derived from the label hierarchy. The hierarchical instance similarity as well as the consolidation result are inferred in a totally unsupervised, iterative fashion. Experimental results on both synthetic and real-world data sets show the effectiveness of the proposed method over existing alternatives.
\end{abstract}



\section{Introduction}
For various tasks such as crowdsourcing, healthcare data mining in big data applications, multiple information sources may provide labeling information on the same instance simultaneously. For example, in crowdsourcing tasks, multiple human annotators are asked to find labels of flowers given a beautiful nature image. On online healthcare forums, a patient who posts a question regarding his/her symptoms may receive disease names as suggestions from multiple doctors. 

Once we obtained the labeling information from multiple information sources or human beings, it is necessary to consolidate the collected information to infer the ground truth labels. Because imperfect information from a single information source exists ubiquitously, it is also important that labeling information from multiple sources need to be consolidated to resolve noises and conflicts. Moreover, due to privacy concerns, raw features of instances are often discarded or withheld and only labels are available for aggregation purposes. For example in online healthcare forums, the raw features of a patient need to be discarded for privacy concerns and only diseases names collected from multiple doctors are consolidated to infer the ground truth.

In those applications, instead of assigning a single label for each instance, it is usually more informative to associate an instance with more than one labels to characterize multiple categories or properties an instance has. For example, an image instance can be described by multiple tags such as ``monarda'', ``bird'' and hence belongs to multiple categories. A protein can be associated with more than one functions, denoting various functionalities. A patient may be associated with several candidate diseases, each of them diagnosed by a doctor. 

\begin{figure}[htb!]
\centering
\epsfig{file=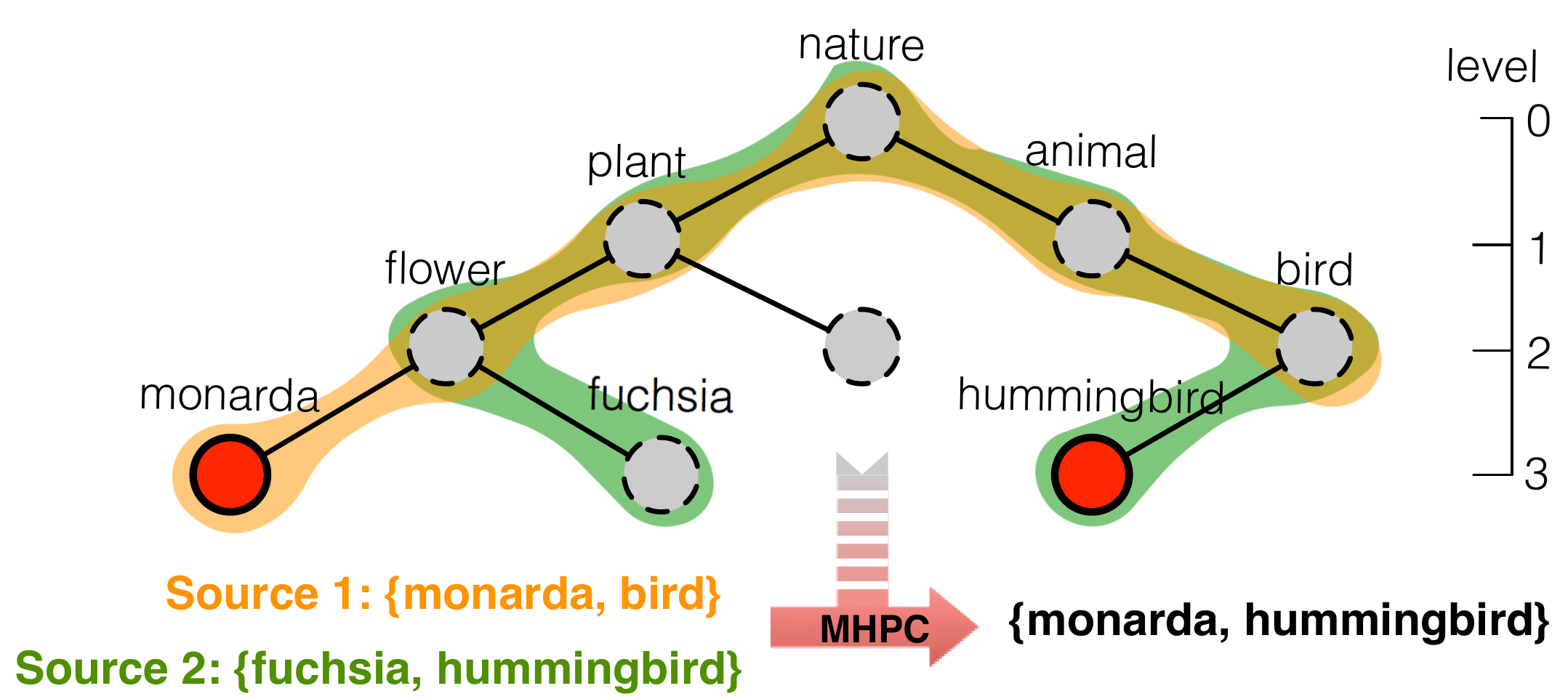,width=3in}
\caption{An illustrative example of multi-source hierarchical prediction consolidation problem. Two individual information sources give label predictions to an image and MHPC tries to find a consolidated label prediction that maximizes the consensus among these predictions while preserving the structures of the label hierarchy.}\label{fig::hierarchical-demo}
\end{figure}

Typically, those tasks consider all the labels on the same ``flat'' level. However, it is still insufficient to measure the value of the informativeness of labels when we isolate labels with each other and ignore the correlations between labels \cite{cheng2012active}. A better way is to organize labels in a hierarchical taxonomy. In this way, besides correlations such as co-occurrences between all the ``flat'' labels, a label hierarchy contains rich information to make sense of highly complex label correlations. 
\\\textbf{Problem Studied}: In this paper, we want to incorporate the label hierarchy into the prediction consolidation process when only the labeling information from multiple information sources are available, which is formally defined as the \underline{M}ulti-source \underline{H}ierarchical \underline{P}rediction \underline{C}onsolidation(MHPC) problem, illustrated in Figure \ref{fig::hierarchical-demo}.

Informative label hierarchies are prevalently observed in various applications. For example, in crowdsourcing for protein functionality annotation, the functional labels of a protein are in a hierarchy, representing the functional relations. In healthcare data mining, disease labels can be also constructed as a tree-like disease taxonomy, denoting the pathology structure of human diseases. 
Current works in the literature \cite{quinlan1996bagging,yu2012transductive, lee2010learning} try to consolidate predictions from multiple information source. However, those works have access to raw features of the data and they totally ignore the hierarchical information. Simply ignore the label hierarchy may lead to potential loss of valuable information \cite{cai2004hierarchical}. 

With a number of approaches proposed to exploit the label hierarchy for various tasks \cite{wang2015exploring,wang2015exploiting} such as classification \cite{bi2012mandatory} or clustering \cite{murtagh2012algorithms}, it is reasonable to believe that informative label hierarchies, which inherently come with many prediction consolidation tasks, are also able to offer auxiliary and valuable information for the MHPC problem. For example, the conflicts between two label predictions in Figure \ref{fig::hierarchical-demo} can be resolved by mapping label predictions to level 2, where we have \{flower, bird\} for both label predictions. Moreover, the label hierarchy may provide some constrains so that label predictions which violate the hierarchical structure will be less useful.

\begin{figure}[tbh!]
\centering
\epsfig{file=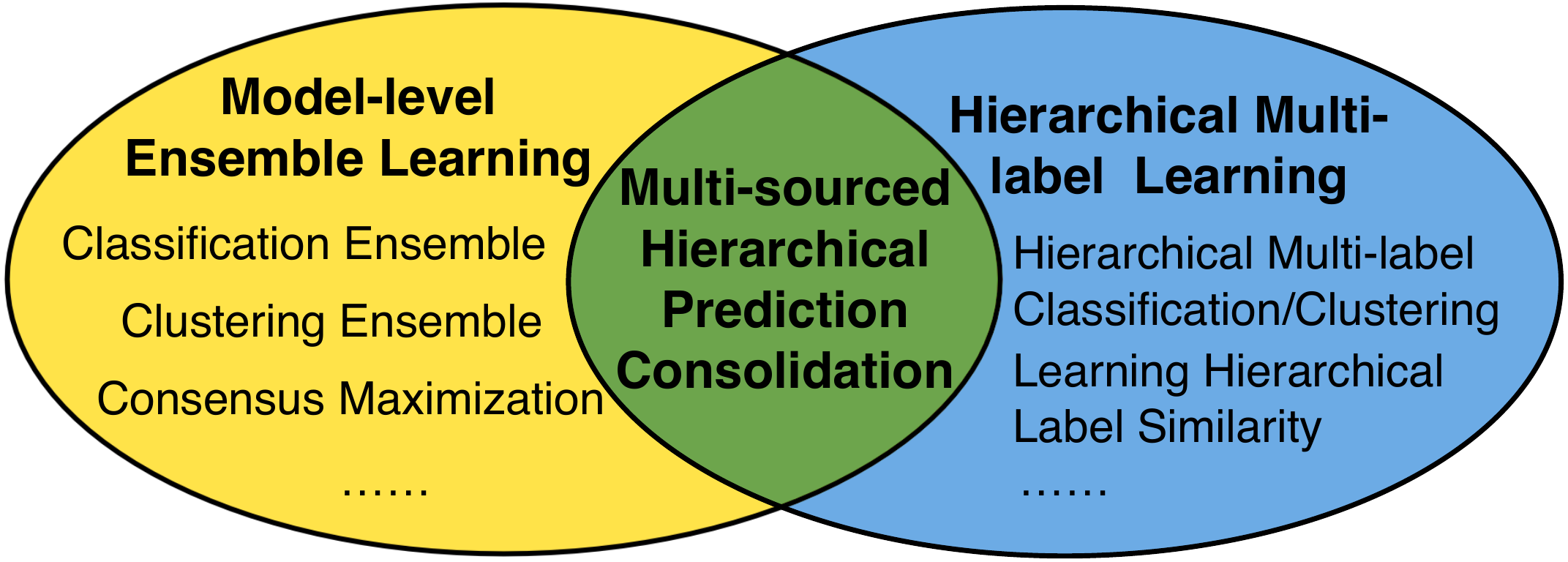,width=3.2in}
\caption{Position of the multi-source hierarchical prediction consolidation problem.}\label{fig::related-works}
\end{figure}

Given the importance of incorporating the label hierarchy, the MHPC problem itself is a novel problem which is rarely studied. Various learning problems are summarized in Figure \ref{fig::related-works}, where model-level ensemble learning \cite{NIPS2009_3855,Xie2013,Dong2015} tries to aggregate labels at the output level and hierarchical multi-label learning exploits the label hierarchy to improve the model performance on a wide range of multi-label learning tasks \cite{cai2004hierarchical,bi2011multi,bi2012mandatory}. The multi-source hierarchical prediction consolidation problem is an unsupervised ensemble learning problem that aggregates hierarchical multi-label predictions on the model-level, where very few work has been done. 

Despite the importance and novelty, the multi-source hierarchical prediction consolidation problems are challenging to solve due to:
\begin{itemize}
\item \textbf{Label vagueness}: 
label vagueness usually originates from imperfect predictive models or  insufficient knowledge of information sources. When an information source has insufficient knowledge or high uncertainty, vagueness is commonly observed where a vague, generalized label is used instead of a mandatory, specific label prediction. 
For example in Figure \ref{fig::hierarchical-demo}, instead of having a specific type of flower (e.g. ``monarda'') as the label prediction, which is the leaf node in a label hierarchy, usually a more generalized label ``flower'' is used when the information source doesn't know much about flowers. Leaf node predictions are more likely to be error-prone when they are mandatory provided under insufficient knowledge. As the label vagueness is widely observed, how can we exploit the rich information encoded within label hierarchy to resolve the vagueness?

\item \textbf{Label ambiguity}: 
in most cases, the predictions we collect from multiple information sources are noisy and conflicting with each other. For instance, in Figure \ref{fig::hierarchical-demo}, Source 2 associates a label ``fuchsia'' to the instance instead of giving ``monarda'', one of the truth labels. Those two labels have similar meanings but may be used interchangeably by different information sources due to ambiguity. Ambiguous label predictions may contain erroneous information and hence introduce noises into prediction consolidation tasks. How should the prediction consolidation task resolve the label ambiguity from multi-source predictions effectively?

\item \textbf{Label sparsity}:
since an information source may provide labeling information for a diverse population of instances, labels in each prediction only cover a very small portion of the whole, diversified label space. Also, many of the labels may be only covering a small number of instances. For example, there can be thousands of flower names under the node ``flower'', but not all information sources necessarily mention the idea of  ``flower'' ever. Not all flower names are covered by all instances as well. 
The worst case exists when the truth label of an instance is not provided by any information source. That is, for the example in Figure \ref{fig::hierarchical-demo}, ``monarda'' is never mentioned by any of the information sources.
 How to deal with the label sparsity that comes with the ever-expanding label space as well as the varying number of predictions obtained from multiple information sources?
\end{itemize}



In this paper, we try to solve the MHPC problem by formulating it as an optimization task. The objective function for optimization favors the smoothness over all information sources as well as penalizing any two instances which have high hierarchical instance similarity but conflict with each other in the consolidation result. We derive a closed-form solution for this optimization problem. After that, the \textsc{Mphc} algorithm is introduced where two phases, namely estimating hierarchical similarities and minimizing consensus cost, are conducted in an iterative, totally unsupervised fashion to get the consolidated label prediction for each instance.

\section{Problem Statement}
Before introducing the proposed method, we will give the definitions of some important concepts and formulation of the MHPC problem first in this section.
\subsection{Terminology Definition}
\begin{table}
	\centering
    \caption{Table of symbols.}\label{tab::symbols}
    \begin{tabular}{c|m{6.5cm}}
	Symbol	& Description				 \tabularnewline  \hline
    $M$  	& Number of information sources \tabularnewline \arrayrulecolor{gray}\hline
	$N$  	& Number of instances           \tabularnewline \hline
	$K$		& Number of labels in the label hierarchy	\tabularnewline \hline
    $\mathbf{Y_m}$		& ${\mathbb{R}}^{N \times K}$ label matrix from source $m$	\tabularnewline \hline
    $\mathbf{\hat Y}$	& ${\mathbb{R}}^{N \times K}$ consolidated label matrix \tabularnewline \hline
    $\mathbf{Y_m}_{(i)}$& Label vector for the $i$-th instance in $Y_m$	\tabularnewline \hline
    $\mathbf{Y_m}_{(i)}^k$& The $k$-th label of $\mathbf{Y_m}_{(i)}$	\tabularnewline \hline
    $\mathbf{Z}$ & ${\mathbb{R}}^{N \times K}$ ground truth label matrix	\tabularnewline \hline
    $\mathbf{H}$& ${\mathbb{R}}^{K \times K}$ hierarchical adjacency matrix of labels where $H_{kk'}=1$ when 	label $k$ is the direct descendant of label ${k'}$. Otherwise $H_{kk'}=0$. \tabularnewline \hline
    $\mathbf{W}$ & $\mathbb{R}^{N \times N}$ hierarchical instance similarity matrix\tabularnewline \hline
    $\vv{\mathbf{S}}$ &  $\mathbb{R}^{1 \times K}$ label support vector.  $\vv{\mathbf{S}}_k$ is the corresponding support value for label $k$.\tabularnewline \hline
    $\vv{\mathbf{C}}$ &  $\mathbb{R}^{1 \times K}$ label occurrence vector.  $\vv{\mathbf{C}}_k$ is the corresponding occurrence value for label $k$ in all augmented label vectors.\tabularnewline \hline
    \end{tabular}
\end{table}

Suppose we are given label matrices $\mathbf{Y}_1$, $\mathbf{Y}_2$,...,$\mathbf{Y}_{M}$ provided by $M$ information sources and a hierarchical adjacency matrix $\mathbf{H}$.
An label matrix $\mathbf{Y}_m$ is an $N$ by $K$ matrix indicating the labeling information on all $N$ instances and $K$ labels provided by the information source $m$. 
Within $\mathbf{Y}_m$, the $i$-th row ${\mathbf{Y}_m}_{(i)}$ is a label vector of instance $i$ provided by information source $m$, where ${\mathbf{Y}_m}_{(i)}^k=1$ if the information source $m$ associates the instance $i$ with label $k$. 
A hierarchical adjacency matrix $\mathbf{H}$ embodies the hierarchical structure for all $K$ labels where $\mathbf{H}_{kk'}=1$ if and only if when $k$ is the direct descendant of label $k'$. Other terminologies are introduced further when they are used. Table \ref{tab::symbols} summarizes the notations.

\subsection{Problem Statement}
Based on the terminologies defined above, the \textbf{Multi-source Hierarchical Prediction Consolidation} problem is formally defined as: given label matrices $\{ {{\mathbf{Y}}_1},{{\mathbf{Y}}_2}, \ldots ,{{\mathbf{Y}}_m}\}$ and a hierarchical adjacency matrix $\mathbf{H}$, the MHPC problem tries to incorporate the label hierarchy into finding a consolidated label matrix $\mathbf{\hat Y}$ which maximizes the consensus among all label matrices. The maximum consensus is achieved by minimizing the consensus cost in prediction consolidation. 

\section{Modeling Hierarchical Consensus}
This section describes how we come up with a consolidation agreement of multiple information sources by incorporating the label hierarchy into the minimization of consensus cost. 
Section \ref{sec::minimizing_consensus_cost} describes the objective function for minimizing the consensus cost as well as its closed-form solution. Section \ref{sec::estimating_hierarchical_similarities} integrates the hierarchical information into the objective function when estimating instance similarities.
\subsection{Minimizing the Consensus Cost}\label{sec::minimizing_consensus_cost}
We formulate the following objective function:
\begin{equation}\label{eq::optimization}
\centering
\begin{gathered}
  \mathop {\min }\limits_{{\mathbf{\hat Y}}} \frac{1}{M}\sum\limits_{m = 1}^M {\left\| {{\mathbf{\hat Y}} - {{\mathbf{Y}}_m}} \right\|_F^2}  + \lambda \sum\limits_{i = 1}^N {\sum\limits_{j = 1}^N {{\mathbf{W}_{ij}}\left\| {{{{\mathbf{\hat Y}}}_{(i)}} - {{{\mathbf{\hat Y}}}_{(j)}}} \right\|_2^2} } \\ \quad s.t. \quad \lambda\geqslant0,
\end{gathered} 
\end{equation}
where ${\left\| {\;.\;} \right\|}$ denotes the $\ell$2-norm for matrices and vectors. The first term favors the smoothness of the consolidation result over label predictions from all the information sources. The second term serves as a regularization term which ensures that label vectors of any two instances in the consolidation result(the $i$-th and the $j$-th instance in $\mathbf{\hat Y}$) do not differentiate themselves from each other very much if they share a high hierarchical instance similarity $\mathbf{W}$. Estimation for $\mathbf{W}$ will be further explained in Section \ref{sec::estimating_hierarchical_similarities}. $\lambda$ serves as a regularization coefficient to penalize violation for the hierarchical similarity constraints. 

Note that the objective function in Equation \ref{eq::optimization} can be rewritten in a matrix form as:
\begin{equation}\label{eq::optimization_matrix}
\centering
\begin{gathered}
\operatorname{J} ({\mathbf{\hat Y}}) = \frac{1}{M}\sum\limits_{m = 1}^M {\left( {\operatorname{Tr} ({{{\mathbf{\hat Y}}}^{\text{T}}}{\mathbf{\hat Y}}) - 2\operatorname{Tr} ({{{\mathbf{Y}}}_m}^{\text{T}}{\mathbf{\hat Y}}) + \operatorname{Tr} ({{{\mathbf{Y}}}_m}^{\text{T}}{{{\mathbf{Y}}}_m})} \right)}  + \\\lambda \operatorname{Tr} ({{{\mathbf{\hat Y}}}^{\text{T}}}{\mathbf{L\hat Y}}) \quad s.t. \quad \lambda\geqslant0,
\end{gathered}
\end{equation}
where $\mathbf{L}$ is the symmetric normalized Laplacian matrix 
\begin{equation}\label{eq::laplacian}
{\mathbf{L}} = {{\mathbf{D}}^{ - \frac{1}{2}}}({\mathbf{D}} - {\mathbf{W}}){{\mathbf{D}}^{ - \frac{1}{2}}}
\end{equation} 
and $\mathbf{D}$ is the degree matrix of $\mathbf{W}$.

To find a $\mathbf{\hat Y}$ that gives the minimum value of $\operatorname{J}(\mathbf{\hat Y})$, we first prove the convexity of $\operatorname{J}(\mathbf{\hat Y})$ by showing the positive definite property of the Hessian matrix of $\operatorname{J}(\mathbf{\hat Y})$ with respect to $\mathbf{{\hat Y}}$.
\begin{equation}\label{eq::hessian}
\frac{{{\partial ^2}\operatorname{J} ({\mathbf{\hat Y}})}}{{\partial {\mathbf{\hat Y}}^2}} = 2\left( {{\mathbf{I}} + \lambda {\mathbf{L}}} \right) \quad s.t. \quad \lambda\geqslant0.
\end{equation}
The Hessian matrix shown in Equation \ref{eq::hessian} is positive definite because $\lambda\mathbf{L}$ is positive-semidefinite \cite{mohar1991laplacian} and adding an identity matrix to it makes the resulting Hessian matrix positive definite \cite{ravindran2006engineering}. Therefore, setting the derivative of $\operatorname{J}(\mathbf{\hat Y})$ with respect to $\mathbf{\hat Y}$ to zero
\begin{equation}\label{eq::first_derivative}
\begin{gathered}
\frac{{\partial \operatorname{J} ({\mathbf{\hat Y}})}}{{\partial {\mathbf{\hat Y}}}} = \frac{1}{M}\sum\limits_{m = 1}^M {\left( {2{\mathbf{\hat Y}} - 2{{{\mathbf{Y}}}_m}} \right)}  + 2\lambda \mathbf{L}{\mathbf{\hat Y}} = 0
\end{gathered}
\end{equation} 
leads to a global minimized consensus cost $\operatorname{J}(\mathbf{\hat Y})$ given $\mathbf{W}$.

A closed-form solution can be obtained by solving the Equation \ref{eq::first_derivative}:
\begin{equation}\label{eq::close_form}
\begin{gathered}
\begin{aligned}
{\mathbf{\hat Y}} &= {\left( {1 + \lambda ({{\mathbf{D}}^{ - \frac{1}{2}}}({\mathbf{D}} - {\mathbf{W}}){{\mathbf{D}}^{ - \frac{1}{2}}})} \right)^{ - 1}}\frac{1}{M}\sum\limits_{m = 1}^M {{{{\mathbf{Y}}}_m}} 
\\
&= {\left( {\mathbf{I} + \lambda  \cdot {\mathbf{L}})} \right)^{ - 1}}{\mathbf{\bar Y}},
\end{aligned}
\end{gathered}
\end{equation}
where $\mathbf{\bar Y}=\frac{1}{M}\sum\limits_{m = 1}^M {{{{\mathbf{Y}}}_m}}$ and $\mathbf{I}$ is the identity matrix.

From Equation \ref{eq::close_form}, we can see that hierarchical similarity constraints are not introduced when $\lambda=0$. In this case, the consolidation result degrades to the simple averaging of all the label prediction we obtained. While $\lambda > 0$, the Laplacian matrix $\mathbf{L}$ regularizes the simple averaging result and guides the $\mathbf{\bar Y}$ towards a global consensus $\mathbf{\hat Y}$ with label hierarchies being considered.

It is worth mentioning that the formulation of the hierarchical similarity constraints as the second term in Equation \ref{eq::optimization} and Equation \ref{eq::optimization_matrix} can be also seen as learning an optimal embedding \cite{belkin2003laplacian} from a multi-source label space to a consolidated label space. 
The multi-source label space contains multi-source hierarchical label predictions with noisy and conflicting labels. While the consolidated label space has labels with less imperfect information as well as a minimized consensus cost. 
If $\mathbf{\hat Y}$ is such an embedding result, then a reasonable criterion for a ``good'' mapping is to have weight $\mathbf{W}_{ij}$ so that it heavily penalizes two ``hierarchically similar'' label predictions ( instances $i$ and $j$ ) for not having ``similar'' label predictions in the consolidated label space after the mapping.

\subsection{Estimating the Hierarchical Similarity}\label{sec::estimating_hierarchical_similarities}
Given any two label vectors, each of which denotes a label prediction for an instance, an instance similarity matrix $\mathbf{W}$ in Equation \ref{eq::optimization} measures similarities between label vectors. We assume that each label has a unique degree of support, asserting that such label belongs to an instance. Then, the instance similarity value $\mathbf{W}_{ij}$ of any two instance $i$ and instance $j$ can be calculated by the following Equation:
\begin{equation}\label{eq::hierarchical_label_weight}
{{\mathbf{W}}_{ij}} = \exp \left( { - \frac{1}{{\sigma }}\sqrt {\sum\limits_{k = 1}^K {{{{\mathbf{\vv S}}}_k}{{({\mathbf{\hat Y}}_{(i)}^k - {\mathbf{\hat Y}}_{(j)}^k)}^2}} } } \right),
\end{equation}
where ${\sqrt {\sum\limits_{k = 1}^K {{{{\mathbf{\vv S}}}_k}{{({\mathbf{\hat Y}}_{(i)}^k - {\mathbf{\hat Y}}_{(j)}^k)}^2}} } }$ in Equation \ref{eq::hierarchical_label_weight} measures the weighted Euclidean distance between label vectors of two consolidated instances in ${\mathbf{\hat Y}}$, namely ${\mathbf{\hat Y}}_{(i)}$ and ${\mathbf{\hat Y}}_{(j)}$, over all $K$ labels.  $\mathbf{\vv S}$ is a support label vector. Each entry $\mathbf{\vv S}_k$ of $\mathbf{\vv S}$ indicates the degree of support of the label $k$ to the distance estimation. Note that the support value $\mathbf{\vv S}_k$ can be seen as how much contribution a label $k$ makes to the overall similarity estimation of an instance having that label. $\sigma$ is a constant factor and the exponential function $\exp\left( . \right)$ converts weighted Euclidean distance measurement to a similarity measurement. 

Usually, we assume that each label has an individual degree of support to the overall similarity estimation. For example, in online healthcare forums, each disease label may have an individual support to a diagnose (each diagnose consists of several disease labels), when we estimate the similarity of two diagnoses. However, as the label space is becoming more and more complicated, such simplified assumption totally ignores correlations among labels and therefore will lead to an inaccurate similarity estimation due to the label sparsity. 
For example, there can be thousands of labels for flower names such as ``monarda'', ``fuchsia'' and so on. However, it is very unlikely that label predictions provided by various information sources cover all of those flower names. Also, different information sources may have preferences in providing certain labels so the remaining labels may be rarely used. Although these labels share the same general idea (the ``flower''), since the support value is calculated on each label separately, none of those labels is able to make any contribution to the support value of label ``flower'' with which they share the general idea. 

The label hierarchy organizes labels in a tree-like structure in which general labels are the ancestors of specific labels. 
With a label hierarchy being incorporated, labels are no longer independent with each other: the general idea is that when an information source assigns the label $k$ to an instance, the existence of that label reflects a direct occurrence of this label to support the instance. Moreover, such label assignment on label $k$ also indicates indirect occurrences, although not explicitly labeled, from its ancestor labels on the label hierarchy. Therefore, we would like to let the occurrence of each label contributes not only to itself, but also to all its ancestor labels in a label hierarchy. 

\begin{figure}[tbh!]
\centering
\epsfig{file=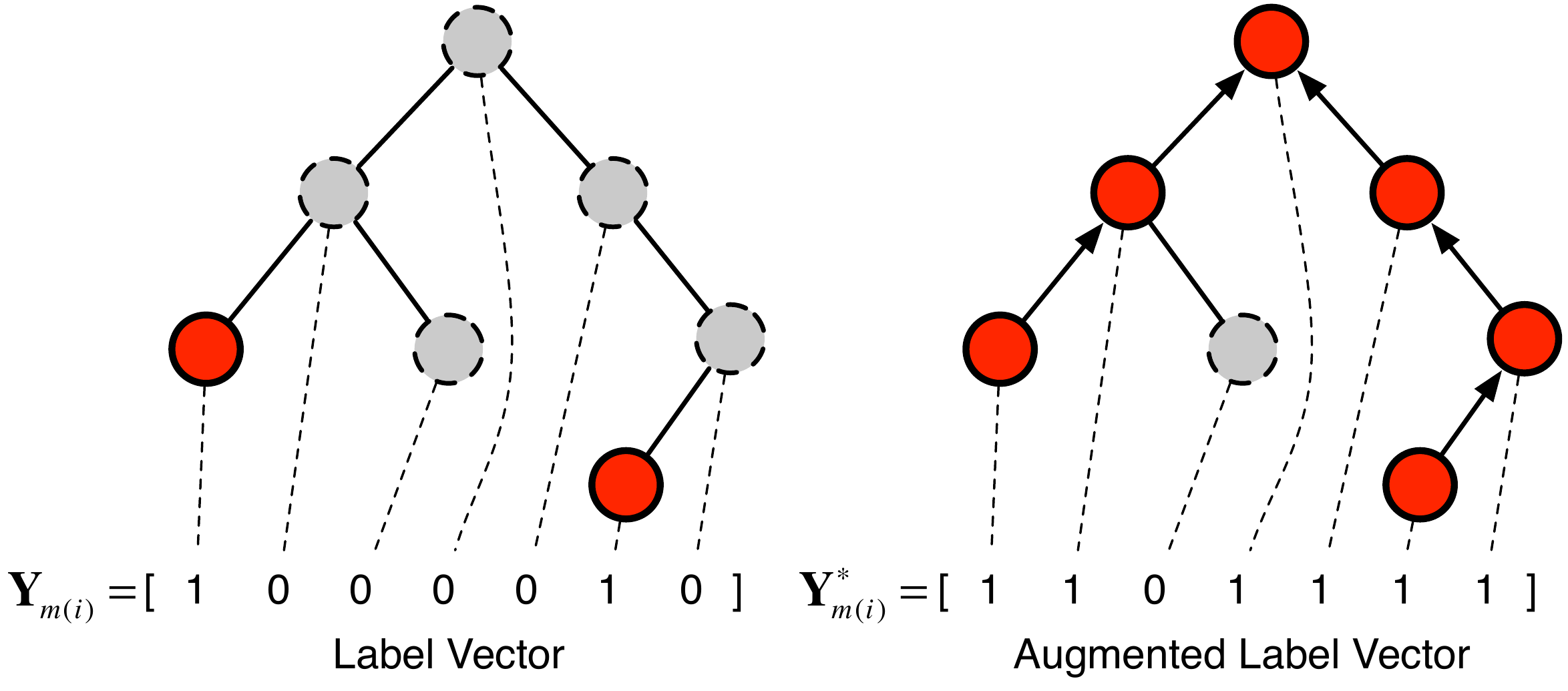,width=3.2in}
\caption{A label augmentation from a label vector $\mathbf{Y_m}_{(i)}$ to an augmented label vector $\mathbf{Y_m}_{(i)}^*$. By augmentation, the direct occurrence of a label in the label vector contributes not only to itself, but also to its ancestor labels in the hierarchy.}\label{fig::label-augment}
\end{figure}

To make this happen, we first apply the label augmentation algorithm to convert each label vector $\mathbf{Y_m}_{(i)}$ to an augmented label vector $\mathbf{Y_m}_{(i)}^*$, as shown in Algorithm \ref{alg::label_augmentation}. By augmenting the occurrence of a label to its ancestor labels, occurrences of those augmented labels provide more implicit, but in-depth labeling information about an instance. The label augmentation for each label vector can be also seen as mapping the original label vector to all the ancestor levels in a bottom-up fashion. Figure \ref{fig::label-augment} illustrates this idea. 
\begin{algorithm}[htb!]
  \caption{Label Augmentation}\label{alg::label_augmentation}
  \begin{algorithmic}[1]
     \Statex \textbf{Input}: A label vector ${{\mathbf{Y}}_{m(i)}}$ 
     \Statex $\quad\quad\quad\,$ A hierarchical adjacency matrix $\mathbf{H}$
	\Statex \textbf{Output}: An augmented label vector ${\mathbf{Y}}_{m(i)}^{k * }$.
    \Statex
    \Function{LabelAugmentation}{${{\mathbf{Y}}_{m(i)}}$, $\mathbf{H}$}
    \State Initialize ${\mathbf{Y}}_{m(i)}^{* }$ as a zero vector
	\For {\textbf{each} $k$ with ${\mathbf{Y}}_{m(i)}^{k }=1$}
    \State $t$ $\gets$ $k$
  	\While  {$t$ $\neq$ \textit{ROOT}} 
    	\State ${\mathbf{Y}}_{m(i)}^{t * } \gets 1$
        \State $t$ $\gets$ $t'$ where $\mathbf{H}_{tt'}=1$
    \EndWhile
    \EndFor\\
    \Return ${\mathbf{Y}}_{m(i)}^{* }$
  \EndFunction
  \end{algorithmic}
\end{algorithm}

We propose to use the occurrence of a label in all augmented label vectors to represent how general information this label embodies, thus the support value for each label can be further estimated. As shown in Equation \ref{eq::summing_up_augmented_label_vector}, by summing up the occurrence of each label in all the augmented label vectors, 
\begin{equation}\label{eq::summing_up_augmented_label_vector}
{{{\mathbf{\vv C}}}_k} = \sum\limits_{m = 1}^M {\sum\limits_{i = 1}^N {{\mathbf{Y}}_{m(i)}^{k * }} }.
\end{equation}
labels close to the root node in the hierarchy are frequently augmented, thus have higher occurrence values ${{{\mathbf{\vv C}}}_k}$. Occurrence values ${{{\mathbf{\vv C}}}}$ are used to calculate support values ${{{\mathbf{\vv S}}}}$ for each label.

Intuitively, the higher occurrence ${{{\mathbf{\vv C}}}_k}$ value the label $k$ gets, the more likely that the label $k$ is near the root node in a label hierarchy. Asserting the belongingness of a label which is in proximity to the root node to an instance is less likely to be fallible. Therefore, labels with high occurrence values makes themselves more supportive in similarity estimation. 

The occurrence value of each label over the sum of occurrences of all labels in all augmented label vectors can be quantified to estimate the support value of each label. However, such estimation can be very inaccurate because although the label predictions are collected from multiple information sources, the occurrences of many labels near the leaf nodes in a hierarchy only share a very small portion over the sum of occurrences, even after label augmentation. This leads to very small values for all the labels. 

In this work, we model the occurrence of a label in a label vector with a confidence interval. The occurrence of a label within all augmented label vectors can be considered as it is sampled from a subset of a population of labels. The occurrence information will be more accurate when we observe this label more frequently among all the augmented label vectors, which leads to a narrow confidence bound. Otherwise, if we rarely observe the occurrence of a label among all label vectors, the resulting confidence bound will be wide and it will be more risky to incorporate this piece of occurrence information into the support value calculation. By incorporating confidence intervals, the occurrence value itself shows how uncertain we are about the occurrence value of a label. 

We use the occurrence value ${{{{\vec C}}_k}}$ of label $k$ over the number of instances $N$ as the proportion. Since the root node will be activated in all the augmented label vectors, the root node will have the highest proportion as 1. One label is either activated or not in an augmented label vector, therefore the binomial probability distribution is used. The confidence interval on this proportion can be presented in an alternate formulation that uses quantiles from the beta distribution \cite{brown2001interval}:
\begin{equation}\label{eq::confidence_interval}
\resizebox{.9\hsize}{!}{$
B\left( {\frac{\alpha }{2};{{{\mathbf{\vv C}}}_k},{N} - {{{\mathbf{\vv C}}}_k} + 1} \right)< \theta_k < B\left( {1-\frac{\alpha }{2};{{{\mathbf{\vv C}}}_k},{N} - {{{\mathbf{\vv C}}}_k} + 1} \right)$},
\end{equation}
where $B$ is the Beta distribution and $\alpha$ is the significance level, which usually has the value 0.05 (5\%). $\theta_k$ is the probability of a label $k$ being activated in an augmented label vector. $\mathbf{\vv C}_{k}$ is the occurrence value of label $k$ over all augmented label vectors.

By Equation \ref{eq::confidence_interval}, we can know that the less frequent a label $k$ being sampled, the wider confidence interval it will end up with. A narrow confidence bound indicates a stronger certainty during the support value calculation. The lower bound value of the confidence interval is used to calculate the support value for each label:
\begin{equation}\label{eq::support_label_vector}
{{{\mathbf{\vv{S}}}}_k} = B\left( {\frac{\alpha }{2};{{{\mathbf{\vv C}}}_k},{N} - {{{\mathbf{\vv C}}}_k} + 1} \right)
\end{equation}
Once we calculate the support value for each label with confidence, the instance similarity matrix $\mathbf{W}$ in Equation \ref{eq::hierarchical_label_weight} can be calculated with the support label vector $\vv{S}$ we learned from the label hierarchy. Hence, $\mathbf{W}$ is called a hierarchical instance similarity matrix.

\section{Consolidating hierarchical labels}
\subsection{The M\textbf{\textsc{HPC}} Algorithm}
Although we provide a closed-form solution to find the global minimum for the consensus cost when the hierarchical instance similarity matrix $\mathbf{W}$ is given, it is still hard to estimate both $\mathbf{\hat Y}$ that associates with a minimized consensus cost $J(\mathbf{\hat Y})$ and the hierarchical instance similarity matrix $\mathbf{W}$ at the same time. Hence, we propose a two-phase iterative algorithm, namely the multi-source hierarchical prediction consolidation (\textsc{Mhpc}) algorithm, that naturally decouples the computation within each iteration.

The \textsc{Mhpc} algorithm has two phases within each iteration, namely estimating the hierarchical similarity (Section \ref{sec::estimating_hierarchical_similarities}) and minimizing the consensus cost (Section \ref{sec::minimizing_consensus_cost}). 
The \textsc{Mhpc} algorithm starts at iteration $t_0$ with an initial estimation for hierarchical instance similarity matrix ${\left( {\mathbf{W}}_{ij} \right)_{{t_0}}}$ and the consolidation result ${\left( \mathbf{\hat Y} \right)_{{t_0}}}$.
${\left( {\mathbf{W}}_{ij} \right)_{{t_0}}}$ is initialized by the following equation:
\begin{equation}
{\left( {\mathbf{W}}_{ij} \right)_{{t_0}}} = \exp \left( { - \frac{1}{{\sigma}}\sqrt {\sum\limits_{k = 1}^K {{{{\mathbf{\vv S}}}_k}{{({\mathbf{\bar Y}}_{(i)}^k - {\mathbf{\bar Y}}_{(j)}^k)}^2}} } } \right),
\end{equation}
where ${\mathbf{\bar Y}}_{(i)}^k $ is the simple averaging result on the $k$-th label of the instance $i$ from all $M$ sources, calculated by = ${\mathbf{\bar Y}}_{(i)}^k = \frac{1}{M}\sum\limits_{m = 1}^M {{{\mathbf{Y}^k}_{m(i)}}} $. 
Note that support values in ${{{\mathbf{\vv S}}}}$ and the occurrence values in ${{{\mathbf{\vv C}}}}$ are derived from all the multi-source label predictions we obtained, which we only initialize once in the entire algorithm, regardless of iterations. 
$({{\mathbf{\hat Y}}})_{t_0}$ is initialized using the Equation \ref{eq::close_form}.

Once we obtain an initial value for ${\left( {{{\mathbf{W}}_{ij}}} \right)_{{t_0}}} $ and $({{\mathbf{\hat Y}}})_{t_0}$, each iteration afterwards follows the following updating rules. \textbf{Estimating the hierarchical similarity}:\\
\begin{equation}
\resizebox{.88\hsize}{!}{$
{\left( {\mathbf{W}}_{ij} \right)_{{t_{x+1}}}} = \exp \left( { - \frac{1}{{\sigma}}\sqrt {\sum\limits_{k = 1}^K {{{{\mathbf{\vv S}}}_k}{{\left( \left( {\mathbf{\hat Y}}_{(i)}^k \right)_{t_{x}} - \left( {\mathbf{\hat Y}}_{(j)}^k \right)_{t_{x}} \right)}^2}} } } \right)
$},
\end{equation}
where the hierarchical instance similarity ${\left( {\mathbf{W}}_{ij} \right)_{{t_{x+1}}}}$ is calculated by the most up-to-date consolidation results in $\left( {{\mathbf{\hat Y}}} \right)_{{t_{{\text{x}}}}}$.\\
\textbf{Minimizing the consensus cost}:\\
\begin{equation}\label{eq::update_consolidation}
{\left( {{\mathbf{\hat Y}}} \right)_{{t_{{\text{x + 1}}}}}} = {\left( {1 + \lambda  \cdot {{\left( {\mathbf{L}} \right)}_{{t_{{\text{x}}}}}}} \right)^{ - 1}}{\left( {{\mathbf{\hat Y}}} \right)_{{t_{{\text{x}}}}}},
\end{equation} where the laplacian matrix $\left({\mathbf{L}}\right)_{t_{\text{x}}} $ is calculated by the most up-to-date $\left({\mathbf{W}}\right)_{t_{x}}$ value using Equation \ref{eq::laplacian}. Note that in Equation \ref{eq::update_consolidation}, the consolidation result in the latest iteration ${( {{\mathbf{\hat Y}}} )_{{t_{{\text{x}}}}}}$ is used for updating, rather than $\mathbf{\bar Y}$ as shown in Equation \ref{eq::close_form}. With this updating function, the consolidation result can accumulate the consolidation progresses from previous iterations. Otherwise, if $(\mathbf{\bar Y})_{t_{x}}$ is used in Equation \ref{eq::update_consolidation}, we simply ignore the consolidation results from all the previous iterations, and $\mathbf{L}_{t_{x}}$ is the only factor we can rely on to guide the consolidation process toward a final consensus. The algorithm terminates whenever the updates on the consolidation result $(\mathbf{\hat Y})_{t_{x+1}}$ is no longer significant after an iteration. 

\section{Experiments}
In this section, we describe the yeast data sets, the real-world medical data sets and their label hierarchies respectively. Experiments on yeast data sets illustrate the ability of the proposed method in overcoming various degree of label vagueness and ambiguity from multiple information sources. While the real-world medical data set emphasizes more on the label sparsity because in real-world medical consultation, we can't ensure that all the information sources provide labels to all instances.

\subsection{Data description and data preprocessing}
\subsubsection{Yeast Data Sets}
Table \ref{tab::yeast_genome} shows statistics about yeast data sets. Each data set annotates yeast genome from a different aspect. Each yeast genome is annotated with hierarchical-structured labels in the Functional Catalogue (FunCat) \cite{ruepp2004funcat}. For example, a yeast genome can be associated with three functionalities:\{20/01/03/01 (sugar transport), 20/03/02/02/01 (proton driven symporter), 20/09/18 (cellular import)\}. The annotation scheme follows the protein functional description of each genome instance, with up to 6 levels of label taxonomy. On average, each instance has 8.8 labels.
\begin{table*}[ht!]
\centering
    \begin{tabular}{lccllllllll}
    \hline
    Data set & seq  & struc & hom   & cellcycle & church & derisi & gasch1 & gasch2 & spo  & expr \\ \hline
\#training & 1701 & 1665  & 1669  & 1628      & 1630   & 1608   & 1634   & 1639   & 1600 & 1639 \\
\#validation & 879  & 860   & 870   & 848       & 844    & 842    & 846    & 849    & 837  & 849  \\ \hline
    \end{tabular}
\caption{Yeast data sets.}\label{tab::yeast_genome}
\end{table*}

Based on the ground truth multi-label label predictions, we introduce vague labels as well as noisy labels to the yeast data sets to model real-world cases where imperfect predictive models or inexperienced human annotators are involved in MHPC problems. 

\begin{figure}[th!]
\centering
\epsfig{file=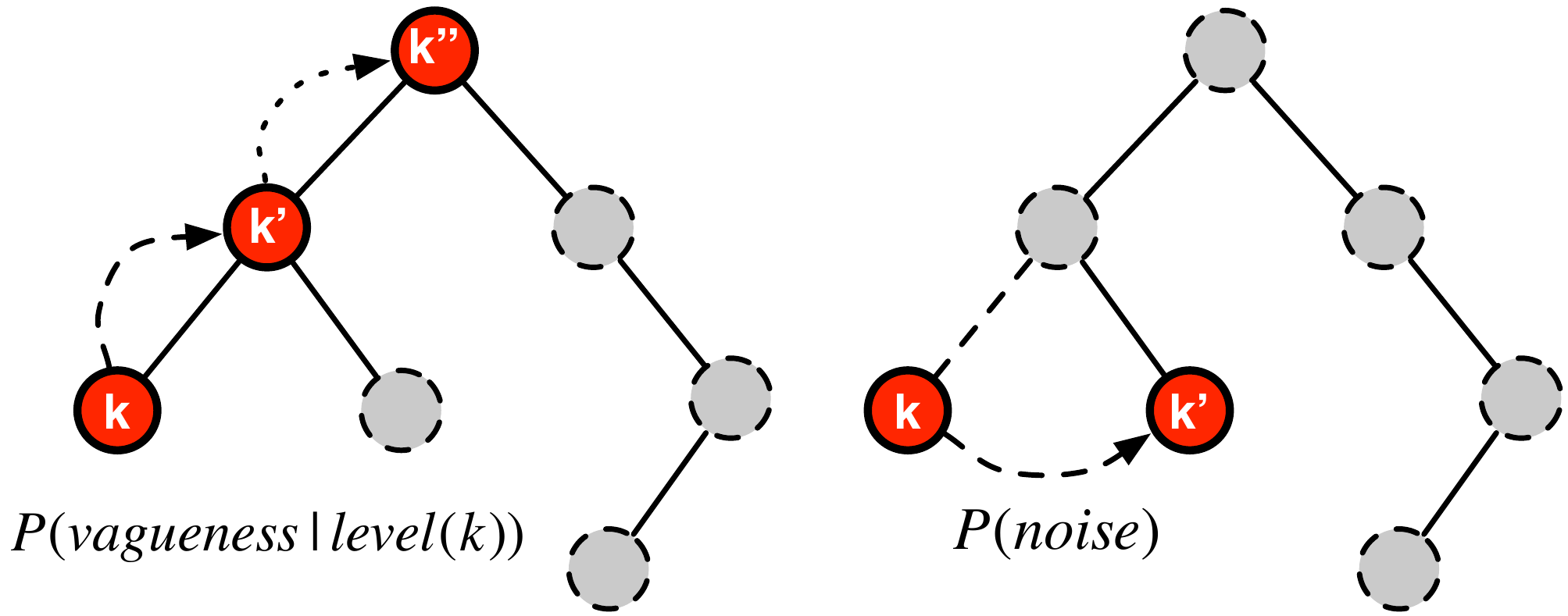,width=3.2in}
\caption{Introducing vague labels and noisy labels to generate the synthetic multi-source label predictions.}\label{fig::missing_noise}
\end{figure}

The Algorithm \ref{alg::generate_labels} is used to generate synthetic label predictions for each instance on all the information sources. For each label mentioned by the ground truth label matrix $\mathbf{Z}$, we generate two random values, $p_V$ and $p_N$ (Line 2-4). Given the label $k$ and a function $level(k)$ indicating which level label $k$ is in, the vagueness of label makes a label $k$ hops to its ancestor $k'$ with a probability $P(vagueness|level(k))$, as shown by the following Equation
\begin{equation}
P(vagueness|level(k)) = {\left( {{\alpha _{vague}}} \right)^{level(k)}},
\end{equation}
where ${\alpha _{vague}}$ is a parameter and $level(k)$ returns the number of connections from label $k$ to the root node. Therefore, a label near the root node of a hierarchy has less probability to hop to its ancestor label, while a label near leaf nodes in the hierarchy is more likely to hop. Whenever $p_V>P(vagueness|level(k))$, we replace $k$ with its ancestor label $k'$. Note that, with one hop performed, the label $k'$ can be further hopped to its ancestor label $k''$ as well (Line 5-9). The left part of Figure \ref{fig::missing_noise} illustrate this idea. 

Once we add the vagueness to the label, when $p_N$ is greater than a transition probability 
\begin{equation}
P(noise) = \alpha_{noise}, 
\end{equation}

noise is introduced by replacing label $k$ by one of its siblings randomly (Line 10-11). Otherwise, the label $k$ will not be changed. The right part of Figure \ref{fig::missing_noise} shows the way we add noise to labels.

\begin{algorithm}[th!]
  \caption{Generating Multi-source Label Predictions}\label{alg::generate_labels}
  \begin{algorithmic}[1]
     \Statex \textbf{Input}: A ground truth label matrix ${\mathbf{Z}}$ 
     \Statex $\quad\quad\quad\,$ A hierarchical adjacency matrix $\mathbf{H}$
	\Statex \textbf{Output}: label matrices $\{{\mathbf{Y}}_{1},{\mathbf{Y}}_{2},...,{\mathbf{Y}}_{M}\}$.
    \Statex
    \Function{GeneratePredictions}{${{\mathbf{Y}}_{m(i)}}$, $\mathbf{H}$}
	\For {\textbf{each} label $k$ of instance $i$ where $\mathbf{Z}_i^k=1$}
        \For {\textbf{each} information source m}
            \State Generate two random values $p_V,p_N \in \left[ 0, 1\right]$.
            \While {k $\neq$ \textit{ROOT}}
              \If{$p_V$>$P(vagueness|level(k))$}
                \State $k$ $\gets$ $k'$ where $\mathbf{H}_{kk'}=1$
              \Else
              	\State Break
              \EndIf
            \EndWhile
            \If {$p_N>P(noise)$}
            	\State $k$ $\in$ $\{k'\,|\, \exists l, \,\mathbf{H}_{k'l}=\mathbf{H}_{kl}\}$
            \EndIf
            \State $\mathbf{Y}_{m(i)}^k \gets 1$ 
        \EndFor
    \EndFor\\
    \Return $\{{\mathbf{Y}}_{1},{\mathbf{Y}}_{2},...,{\mathbf{Y}}_{M}\}$
  \EndFunction
  \end{algorithmic}
\end{algorithm}

\subsubsection{Medical Data Sets}
The medical data set and the disease label hierarchy are obtained from an online medical consultation website xywy.com\footnote{http://club.xyxy.com}, where patients post their healthcare related questions and multiple medical professionals give online suggestions or general advice as answers. 

\begin{table*}[thb]
\centering
\begin{tabular}{cccc}
\hline
Instance & Doctor\_id & Labels & Ground Truth   \\ \hline
\multirow{3}{*}{\parbox{3cm}{sneezing, runny nose, sleepy}} & 52****17 & \{common cold, allergic rhinitis\} & \multirow{ 3 }{*}{\parbox{4cm}{\{common cold, allergic rhinitis, rhinitis, sinusitis, antritis\}}}  \\
& 46****35 & \{common cold, rhinitis\} &   \\
& 53****11 & \{rhinitis, sinusitis\} & \\ \hline
\end{tabular}
\caption{Each instance is a set of symptoms that a user describes. When an instance describes a set of symptoms, disease names are the labels we collected from different doctors. Each doctor is considered as an information source that provides disease names as labels.}\label{tab::medical_example}
\end{table*} 
Table \ref{tab::medical_example} gives an example of the multi-source label predictions we collect on an instance. Each doctor is considered as an individual information source. Ground truth disease labels are obtained by a medical knowledge base in Baidu Baike\footnote{http://baike.baidu.com}(an online encyclopedia of Baidu) where registered doctors provide knowledge about certain disease names that closely associate with some symptoms. 
The disease label hierarchy is organized in an anatomical structure, with up to three levels of labels(e.g. disease - otorhinolaryngology - rhinitis). In terms of the label sparsity, 0.011506\% of labels are activated among label predictions over all instances from all information sources. Such label low sparsity is common in data sets like this because not all doctors provide labels for every instance; not all instances get label predictions from each doctor. Usually a doctor provides around 2.5 labels to an instance on average, which leads to a low label coverage over all labels in a hierarchy.


\subsection{Experiment Settings}
\subsubsection{Comparison Methods}
To show the advantages of the \textsc{Mhpc} algorithm in solving multi-source hierarchical prediction consolidation problem, we compare the \textsc{Mhpc} method with many baseline methods. Considering that no known multi-source hierarchical prediction consolidation methods are available, averaging methods as well as other model-level ensemble learning methods are introduced, which can be divided into three categories: \\
\textbf{Averaging Methods}\\
\begin{itemize}
\vspace{-0.2in}
\item \textsc{Sa}: The simple averaging method. The \textsc{Sa} method simply takes the average of multi-source label predictions. In \cite{dani2006empirical}, the authors observe that the simple averaging method is competitive with a variety of adaptive algorithms under the quadratic loss criterion.
\vspace{-0.1in}\item \textsc{Wa}: The weighted averaging method. Besides the simple averaging method which considers an equal contribution of each label to the consolidation result, each label has a support value as a weight learned from Section \ref{sec::estimating_hierarchical_similarities}.
\end{itemize}
\textbf{Consensus Maximization Methods}\\
\begin{itemize}
\vspace{-0.2in}
\item \textsc{Mlcm}: The multi-label consensus maximization method is introduced in \cite{Xie2013}. The \textsc{Mlcm} learns a consolidation result from both label predictions and cluster predictions of the same instance from multiple information sources. By ignoring the cluster predictions which assign each instance with a cluster id, the \textsc{Mlcm} adapts to the problem setting of \textsc{Mhpc}. Note that, the label hierarchy is not explored in \textsc{Mlcm}. The \textsc{Mlcm} formulates a bipartite graph where instance nodes are on one side, label nodes from multiple information source are on the other side. The algorithm learns a subset of the connections between two partitions of nodes while maximizing the consensus among them.
\end{itemize}
\textbf{Multi-source Prediction Consolidation Methods}\\
\begin{itemize}
\vspace{-0.2in}
\item  \textsc{Mpc-u}: The mutli-source prediction aggregation method that minimizes the consensus cost as mentioned in Section \ref{sec::minimizing_consensus_cost}, but uses a uniformed value for each entry of the support label vector $\vv S$ during instance similarity estimation.
\vspace{-0.1in}\item \textsc{Mhpc}: The proposed method which minimizes the consensus cost by optimization and incorporates the label hierarchy in estimating hierarchical instance similarities. The support value of each label is estimated based on the lower bound confidence interval of the proportion of occurrence based on all the augmented label vectors we obtained.
\end{itemize}
\begin{table}[]
\centering
\resizebox{\columnwidth}{!}{%
\begin{tabular}{ccccc}
\hline
      & \multicolumn{2}{l}{Minimizing Consensus Cost} & \multicolumn{2}{c}{Label Weights}           \\ \cmidrule(r){2-3}\cmidrule(l){4-5}
      & Averaging            & Optimizaiton           & Uniform  & Hierarchical \\ \hline
\textsc{Sa}    & \checkmark           &                        & \checkmark       &                               \\
\textsc{Wa}    & \checkmark           &                        &                            & \checkmark            \\
\textsc{Mlcm}  &                    & \checkmark            &         &                             \\
\textsc{Mpc-u} 		   &                      & \checkmark          & \checkmark       &                                 \\
\textsc{Mhpc}  &                      & \checkmark             &           & \checkmark           \\\hline
\end{tabular} }
\caption{Comparison Methods}
\end{table}

\subsubsection{Evaluation metrics}
Ranking loss, micro-AUC and coverage error are three metrics that we used for performance evaluation. 

Ranking loss averages over the instances to penalize the number of label pairs within each instance that are incorrectly ordered. Since the label space is large, it is relatively hard to assign a precise probability to each label from a large label space. Ranking labels become an alternative, sometimes a must, for the evaluation. Perfect ordered labels in instances have zero ranking losses. 

AUC (Area Under the Curve) is designed for binary classification problems with skew class distributions. In hierarchical label predictions, the ground truth labels of an instance are relevant labels that covers a very small portion of the label space. The ground truth labels are dominated by other irrelevant labels. In such scenario, AUC is adopted as a metric which compares the ranks of all possible pairs of labels in terms of the relevance. Formally, the label matrix $\mathbf{\hat Y}$ has a total of $N \times K$ entries. Let $Pos$ be the label set with positive (relevant) entries and $Neg$ be the label set with all the other negative (irrelevant) entries. In hierarchical label predictions we usually have $card(Pos) \ll card(Neg)$, where $card(.)$ is the carnality of a set. Given a list of relevance scores $f(·)$ of all entries, micro-AUC \cite{cortes2004auc} is defined as
\begin{equation}
\operatorname{micro-AUC}  = \sum\limits_{i \in Pos} {\sum\limits_{j \in Neg} {\frac{{\mathbbm{1}\left[ {f(i) > f(j)} \right]}}{{\operatorname{card} (Pos) \times \operatorname{card} (Neg)}}} },
\end{equation}
where $\operatorname{f}(i)$ is the relevance score for entry $i$ and $\mathbbm{1}$ is the indicator function.

Note that what micro-AUC differs from ranking loss is that micro-AUC compares the ranks of any pair of labels, whether those two labels are from the same instance or not. While the ranking loss focuses on the label ranking of individual instances. That is, the difference of ranks between labels of two different instances are not explored.

Since label predictions can be anywhere on the label hierarchy, coverage error \cite{tsoumakas2009mining} is adopted to measure the average number of labels that have to be chosen from the consolidation result so that those labels are able to cover all the ground truth labels.


\subsection{Experimental results}

\subsubsection{Convergence Analysis}
\begin{figure*}[tbh!]
\centering
\epsfig{file=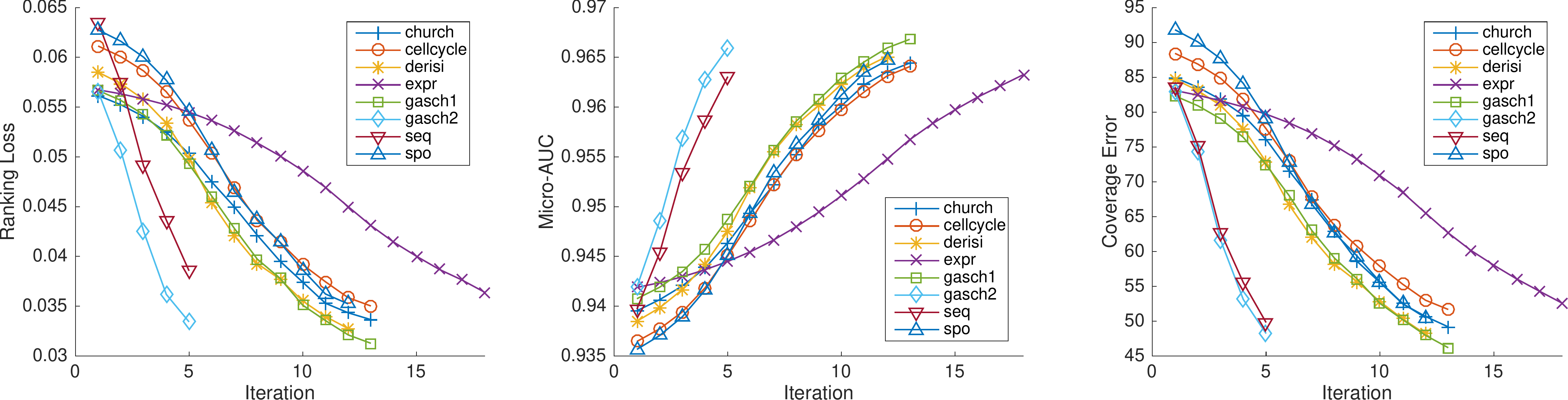,width=7in}
\caption{Ranking loss, mirco-AUC and coverage error on all the yeast data sets as the updatings are conducted iteratively in MHPC.}\label{fig::converge}
\end{figure*}
In the \textsc{Mhpc} algorithm, the hierarchical instance similarity matrix $\mathbf{W}$ and the consolidation result $\mathbf{\hat Y}$ are updated by two phases, namely minimizing the consensus cost and estimating the hierarchical similarity, respectively. Two phases are performed iteratively until convergence. To show that with proper parameters learned form the validation data sets, the two-phase updating rules can lead to a convergence, we show the performance of the \textsc{Mhpc} algorithm after each iteration, as shown in Figure \ref{fig::converge}. Note that, for each data set, the parameter $\lambda$ and $\sigma$ are learned by the validation set. Also, $\alpha_{vague}$ is fixed to 0.8 and $\alpha_{noise}$ is fixed to 0.5 for all data sets. Multi-source label predictions from four information sources are incorporated.

As shown in the figures, as the two-phase updating continues, three evaluation metrics are consistently ameliorated.


\subsubsection{Parameter Estimation}
The parameters of the \textsc{Mhpc} method is chosen by those parameters who give the best performance of the multi-source hierarchical prediction consolidation task on the validation set. The validation set is a portion of the original data sets for parameter learning. 

\begin{figure*}[thb!]
\centering
\epsfig{file=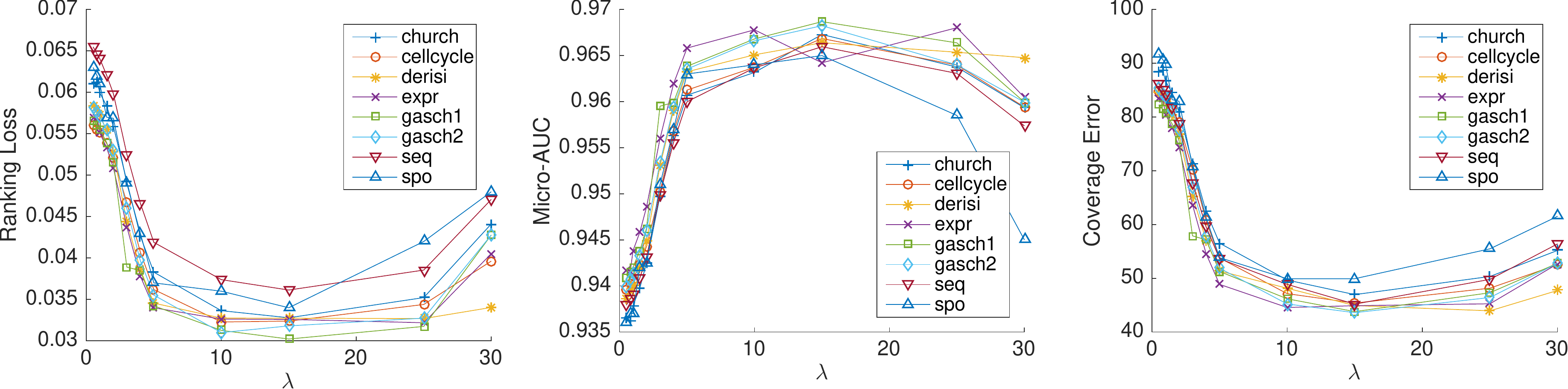,width=7in}
\caption{Parameter estimation for $\lambda$ on yeast data sets.}\label{fig::para_est_yeast}
\end{figure*}

We compare the performance of the \textsc{Mhpc} method on four information sources on all the yeast data sets, with $\alpha_{vauge}$ and $\alpha_{noise}$ fixed as 0.8 and 0.5. Figure \ref{fig::para_est_yeast} illustrates the impact of the value of $lamda$, as a parameter, to the overall performance of the proposed method on the validation set. 

After parameter learning, $\lambda=10$ is chosen for church, cellcyle, derisi and gasch2 data sets; $\lambda=25$ for the expr data set; $\lambda=15$ for gasch1, seq and spo data sets. For medical date sets, we did the same analysis and all the data sets performs the best with $\lambda=26$.

\subsubsection{Sensitivity Analysis}
For experiments on yeast data sets above, two parameters ($\alpha_{vague}$ and $\alpha_{noise}$) used to generate multi-source label predictions are set as fixed values. In this section, we further varies both $\alpha_{vague}$ and $\alpha_{noise}$ from 0.1 to 1 to test how sensitive vague and noisy labels will affect the model performance. We compare the performance of \textsc{Mhpc} method with other alternatives on each combination of $\alpha_{vague}$ and $\alpha_{noise}$. Due to space limitations, only the result on church, one of the yeast data set, is reported with $\lambda=10$. Multi-source label predictions from four information sources are generated.
\begin{figure*}[bht!]
\centering
\epsfig{file=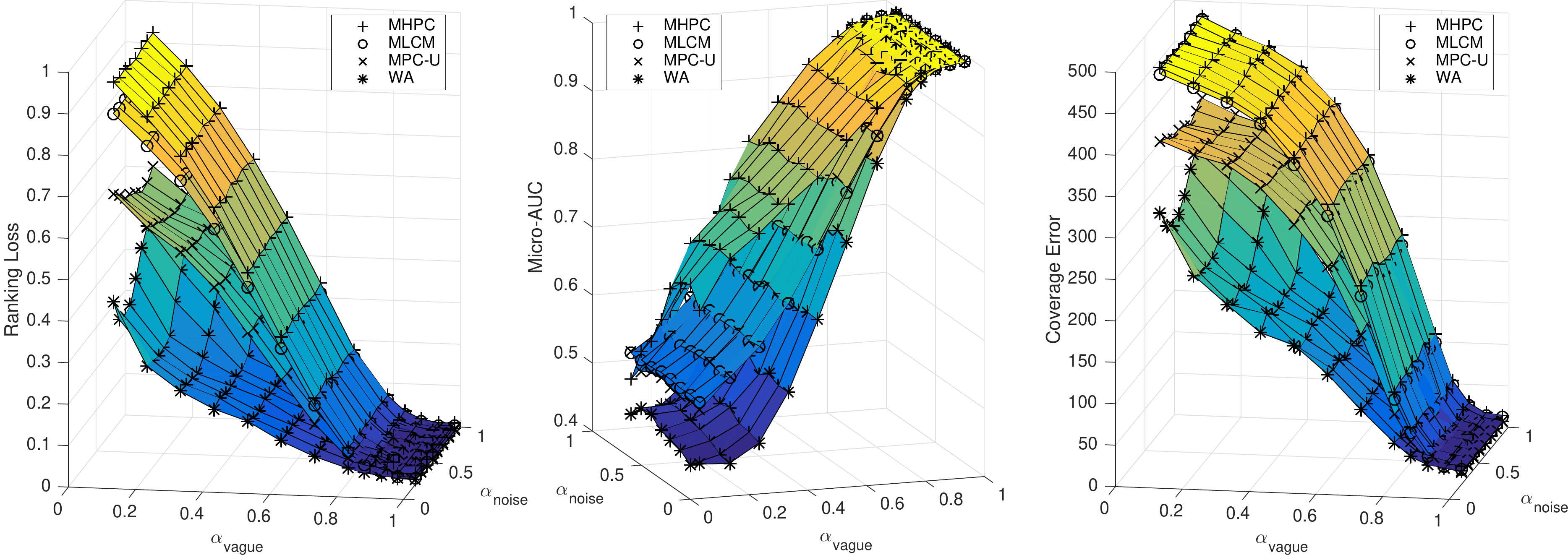,width=7in}
\caption{Performance with varying degrees of vagueness and noises on church, one of the yeast data sets. }\label{fig::task3_all}
\end{figure*}

Figure \ref{fig::task3_all} shows the performance comparisons with three evaluation metrics. We observe that the MPHC outperforms other alternatives consistently. When the vagueness level $\alpha_{noise}$ increases, the performance deteriorates for all the methods. But the \textsc{Mhpc} performs relatively better than others. On the other hand, when the noise level increases and $\alpha_{vague}$ is fixed, \textsc{Wa} and \textsc{Mpc-u} methods are more easily affected by the noisy label, which leads to fluctuations of the performance surfaces. While \textsc{Mhpc} has a relative stable performance when we varies $\alpha_{noise}$ from 0.1 to 1 on almost all the values $\alpha_{vauge}$ can take. Note that the performance of \textsc{Sa} is similar with \textsc{Wa}, so the performance of \textsc{Sa} is not presented in this figure.\\

\subsubsection{Varying the Number of Information Sources}
We vary the number of information sources that we collect the multi-source label predictions from. Evaluation results on all three metrics with cellecycle, one of the yeast data sets is presented in Table \ref{tab::num_of_source}. $\alpha_{vauge}=0.8$ and $\alpha_{vauge}=0.5$ are used to generate multi-source label prediction and $\lambda=10$ is used as the parameter.
\begin{table}[]
\centering
\resizebox{\columnwidth}{!}{%
\begin{tabular}{cccc}
\hline
\multicolumn{1}{l}{Information} & \multirow{2}{*}{Ranking Loss} & \multicolumn{1}{l}{\multirow{2}{*}{Micro-AUC}} & \multicolumn{1}{l}{\multirow{2}{*}{Coverage Error}} \\
\multicolumn{1}{l}{Source}      &                               & \multicolumn{1}{l}{}                           & \multicolumn{1}{l}{} \\\hline
1                                      & 0.171  & 0.830             & 180.679                   \\
3                                      & 0.085 & 0.920             & 102.949                  \\
4                                      & 0.064 & 0.941             & 77.910                  \\
5                                      & 0.054 & 0.952             & 64.182                   \\
10                                     & 0.046 & 0.964             & 47.717                   \\
15                                     & 0.042 & 0.967             & 44.019                   \\
20                                     & 0.042 & 0.967             & 44.041                  \\
30                                     & 0.042 & 0.967             & 43.901                   \\
50                                     & 0.041 & 0.968             & 43.592	\\\hline                  
\end{tabular}
}
\caption{Performance of the MHPC method with label predictions collected from a varying number of information sources.}\label{tab::num_of_source}
\end{table}
We vary the number of information sources from 1 to 50. From Figure \ref{tab::num_of_source} we can see that if we assume information sources are making errors independently, then collecting label predictions from three information sources will cut off almost 50\% of the ranking loss and coverage error, when comparing with label prediction form a single information source. Moreover, as we collect information from more information sources, three evaluation metrics tend to stabilize to a final value. When adding an information source leads to an extra cost (pay a human annotator for labeling), this result gives some insights about the trade-off between the number of information sources we collect labels form, with the extra performance improvement we may get, based on the independent assumption.

\subsubsection{Resolving the Label Sparsity on Medical Data Sets}
The medical data sets come with multi-source label predictions inherently when multiple doctors gives suggestions to each patient. Since the medical date sets have six subsets (MED.1 to MED.6), we randomly sampled 500 instances from each data set. The label predictions associate with those instances are used for prediction consolidation. Table \ref{tab::perfoamance_medical} shows the performance of \textsc{Mhpc} with other baseline methods, for which we can see the superior performance of \textsc{Mhpc} method in real-world scenarios.
\begin{table}[h!]
\centering
\resizebox{\columnwidth}{!}{%
\begin{tabular}{llccc}
\hline
\multirow{2}{*}{Datasets} & \multirow{2}{*}{Methods} & \multicolumn{3}{c}{Evaluation Metrics}    \\\cmidrule{3-5}
                          &                & Ranking Loss & Micro-AUC & Coverage Error \\\hline
\multirow{5}{*}{MED.1}    &\textsc{Sa}& 0.520 {\color{blue}(5)} & 0.620 {\color{blue}(5)} & 213.175 {\color{blue}(5)} \\
                          &\textsc{Wa}& 0.518 {\color{blue}(4)} & 0.621 {\color{blue}(3)} & 212.844 {\color{blue}(4)}\\
                          &\textsc{Mpc-u}& 0.304 {\color{blue}(3)} & 0.547 {\color{blue}(5)} & 125.108 {\color{blue}(3)}\\
                          &\textsc{Mlcm}& 0.262 {\color{blue}(2)} & 0.643 {\color{blue}(2)} & 108.146 {\color{blue}(1)}\\
                          &\textsc{Mhpc}& 0.196 {\color{blue}(1)} & 0.754 {\color{blue}(1)} & 125.107 {\color{blue}(2)}\\\hline
\multirow{5}{*}{MED.2}    &\textsc{Sa}& 0.358 {\color{blue}(5)} & 0.603 {\color{blue}(4)} & 75.544 {\color{blue}(5)} \\
                          &\textsc{Wa}& 0.357 {\color{blue}(4)} & 0.604 {\color{blue}(3)} & 75.416 {\color{blue}(4)} \\
                          &\textsc{Mpc-u}& 0.200 {\color{blue}(2)} & 0.539 {\color{blue}(5)} & 42.630 {\color{blue}(3)} \\
                          &\textsc{Mlcm}& 0.268 {\color{blue}(3)} & 0.680 {\color{blue}(2)} & 31.221 {\color{blue}(1)} \\
                          &\textsc{Mhpc}& 0.166 {\color{blue}(1)} & 0.715 {\color{blue}(1)} & 35.336 {\color{blue}(2)}\\\hline
\multirow{5}{*}{MED.3}    &\textsc{Sa}& 0.067 {\color{blue}(4)} & 0.706 {\color{blue}(2)} & 3.400  {\color{blue}(4)} \\
                          &\textsc{Wa}& 0.065 {\color{blue}(3)} & 0.704 {\color{blue}(3)} & 3.314  {\color{blue}(3)} \\
                          &\textsc{Mpc-u}& 0.069 {\color{blue}(5)} & 0.432 {\color{blue}(5)} & 3.543 {\color{blue}(5)} \\
                          &\textsc{Mlcm}& 0.064 {\color{blue}(2)} & 0.542 {\color{blue}(4)} & 3.288 {\color{blue}(2)} \\
                          &\textsc{Mhpc}& 0.038 {\color{blue}(1)} & 0.847 {\color{blue}(1)} & 2.000  {\color{blue}(1)} \\\hline
\multirow{5}{*}{MED.4}    &\textsc{Sa}& 0.301 {\color{blue}(5)} & 0.601 {\color{blue}(4)} & 38.325 {\color{blue}(5)} \\
                          &\textsc{Wa}& 0.300 {\color{blue}(4)} & 0.602 {\color{blue}(3)} & 38.275 {\color{blue}(4)} \\
                          &\textsc{Mpc-u}& 0.173 {\color{blue}(2)} & 0.460 {\color{blue}(5)} & 22.242 {\color{blue}(3)} \\
                          &\textsc{Mlcm}& 0.225 {\color{blue}(3)} & 0.617 {\color{blue}(2)} & 19.325 {\color{blue}(2)} \\
                          &\textsc{Mhpc}& 0.118 {\color{blue}(1)} & 0.707 {\color{blue}(1)} & 15.233 {\color{blue}(1)} \\\hline
\multirow{5}{*}{MED.5}    &\textsc{Sa}& 0.355 {\color{blue}(5)} & 0.563 {\color{blue}(4)} & 67.534 {\color{blue}(5)} \\
                          &\textsc{Wa}& 0.353 {\color{blue}(4)} & 0.564 {\color{blue}(3)} & 67.170 {\color{blue}(4)} \\
                          &\textsc{Mpc-u}& 0.189 {\color{blue}(1)} & 0.549 {\color{blue}(5)} & 36.080 {\color{blue}(2)} \\
                          &\textsc{Mlcm}& 0.276 {\color{blue}(3)} & 0.565 {\color{blue}(2)} & 52.523 {\color{blue}(3)} \\
                          &\textsc{Mhpc}& 0.232 {\color{blue}(2)}  & 0.572 {\color{blue}(1)} & 32.648 {\color{blue}(1)} \\\hline
\multirow{5}{*}{MED.7}    &\textsc{Sa}& 0.365 {\color{blue}(5)} & 0.598 {\color{blue}(3)} & 47.175 {\color{blue}(5)} \\
                          &\textsc{Wa}& 0.363 {\color{blue}(4)} & 0.599 {\color{blue}(2)} & 46.991 {\color{blue}(4)} \\
                          &\textsc{Mpc-u}& 0.236 {\color{blue}(3)} & 0.491 {\color{blue}(5)} & 30.632 {\color{blue}(3)} \\
                          &\textsc{Mlcm}& 0.190 {\color{blue}(1)} & 0.632 {\color{blue}(1)} & 24.798 {\color{blue}(2)} \\
                          &\textsc{Mhpc}& 0.203 {\color{blue}(2)} & 0.594 {\color{blue}(4)} & 22.633  {\color{blue}(1)} \\\hline

\end{tabular}
}
\caption{Performance on medical data sets.}\label{tab::perfoamance_medical}
\end{table}

\section{Conclusions}
As information explodes, we are able to obtain an increasing number of label predictions from a large population of information sources at the same time. Due to privacy concerns or storage limitations, the raw features of instances are usually discarded or withheld. The labels we collect from multiple information sources bring not only diversity of labels, but also vagueness and noises. In this work, we studied the multi-source hierarchical prediction consolidation (MHPC) problem. Traditional model-level ensemble learning problems deal with multi-source information but they simply ignore the hierarchical structure of labels. On the other hand, hierarchical multi-label learning problems try to bring the label hierarchy into varies tasks such as classification. The MHPC problem effectively incorporates the existing label hierarchy to resolve vagueness and noise originate from multiple information sources, where very few works have been done. We formulate the MHPC problem as an optimization task with a closed-form solution. Two phases, namely minimizing the consensus cost and estimating the hierarchical similarity, are performed in an iterative fashion to learn a consolidation result while preserving the structures of the label hierarchy. Experiments conducted on both synthetic and real-world data sets show the advantages of the proposed method over other alternatives.

\bibliographystyle{unsrt}
\bibliography{sigproc}  
\balancecolumns
\end{document}